\documentclass[twocolumn,showpacs,showkeys,amsmath,amssymb,superscriptaddress,floatfix,nofootinbib]{revtex4}

\usepackage{graphicx}
\usepackage{bm} 
\usepackage{subfigure}
\usepackage[T1]{fontenc}

\def\vec#1{\mathchoice{\mbox{\boldmath$\displaystyle#1$}}
{\mbox{\boldmath$\textstyle#1$}}
{\mbox{\boldmath$\scriptstyle#1$}}
{\mbox{\boldmath$\scriptscriptstyle#1$}}}
\makeatletter
\newcommand\erfc{\mathop{\operator@font erfc}\nolimits}
\def\slashchar#1{\setbox0=\hbox{$#1$}
   \dimen0=\wd0 \setbox1=\hbox{/} \dimen1=\wd1
   \ifdim\dimen0>\dimen1 \rlap{\hbox to \dimen0{\hfil/\hfil}} #1
   \else  \rlap{\hbox to \dimen1{\hfil$#1$\hfil}} / \fi}

\begin{document}
 
\title{Soft heavy-ion physics from hydrodynamics with statistical hadronization \\-- predictions for the Large Hadron Collider%
\footnote{Partly supported by the Polish Ministry of Science and Higher Education, grants N202 153 32/4247 and N202 034 32/0918.}}

\author{Mikolaj Chojnacki}
\email{Mikolaj.Chojnacki@ifj.edu.pl}
\affiliation{The H. Niewodnicza\'nski Institute of Nuclear Physics, Polish Academy of Sciences, PL-31342 Krak\'ow, Poland}

\author{Wojciech Florkowski} 
\email{Wojciech.Florkowski@ifj.edu.pl}
\affiliation{The H. Niewodnicza\'nski Institute of Nuclear Physics, Polish Academy of Sciences, PL-31342 Krak\'ow, Poland}
\affiliation{Institute of Physics, \'Swi\c{e}tokrzyska Academy, ul.~\'Swi\c{e}tokrzyska 15, PL-25406~Kielce, Poland} 

\author{Wojciech Broniowski} 
\email{Wojciech.Broniowski@ifj.edu.pl} 
\affiliation{The H. Niewodnicza\'nski Institute of Nuclear Physics, Polish Academy of Sciences, PL-31342 Krak\'ow, Poland}
\affiliation{Institute of Physics, \'Swi\c{e}tokrzyska Academy, ul.~\'Swi\c{e}tokrzyska 15, PL-25406~Kielce, Poland} 

\author{Adam Kisiel} 
\email{kisiel@if.pw.edu.pl}
\affiliation{Faculty of Physics, Warsaw University of Technology, PL-00661 Warsaw, Poland}
\affiliation{Department of Physics, Ohio State University, 1040 Physics Research Building, 191 West Woodruff Ave., Columbus, OH 43210, USA}

\date{6 December 2007}

\begin{abstract}
Hydrodynamics merged with single-freeze-out statistical hadronization is used to describe the midrapidity hadron production 
in relativistic heavy-ion collisions
at the highest RHIC energies ($\sqrt{s_{NN}}=200~{\rm GeV}$) and to make predictions for the LHC ($\sqrt{s_{NN}}=5.5~{\rm TeV}$). Thermodynamic properties of the high-temperature strongly-interacting quark-gluon plasma are taken from  lattice simulations, at low temperatures the hadron-gas equation of state is used, while in the cross-over region an interpolation between the two equations of state is constructed.  Boost invariance is assumed for the midrapidity calculations. The initial condition for hydrodynamics is obtained from a Glauber profile for the entropy, with the initial central temperature $T_i$. The  conditions obtained from the hydrodynamic expansion at the freeze-out temperature $T_f$ are used as input for the thermal event generator {\tt THERMINATOR}, which accounts for a complete treatment of hadronic resonances. Basic physical observables are obtained: the transverse-momentum spectra, the elliptic flow coefficient $v_2$, and the HBT radii. The femtoscopic observables are evaluated with the help of the two-particle method which accounts for the resonance decays and Coulomb final-state interactions. The problem of a simultaneous description of all discussed observables is addressed, with the conclusion that at the highest RHIC energies our approach gives a quite satisfactory global description of soft hadronic observables, which agree with the data at the level of 10-15\%. Some discrepancies may be attributed to the absence of the final-state elastic interactions among hadrons. Extrapolating $T_i$ to higher values allows for global predictions for soft hadronic physics at the LHC. We test $T_i=400$, $450$, and $500$~MeV, and observe the expected growth of particle multiplicities and the increase of the flow, resulting in smaller slopes of the $p_T$-spectra. The elliptic flow of pions exhibits saturation, with $v_2$ remaining practically constant, while the HBT radii increase moderately with $T_i$.   
\end{abstract}



\maketitle 

\section{Introduction}
\label{sect:intro}

Hydrodynamics has become the standard framework for the description of the intermediate stages of relativistic heavy-ion collisions \cite{Kolb:2003dz,Huovinen:2003fa,Shuryak:2004cy,Muller:2007rs,Nonaka:2007nn}. In this paper, encouraged by the success of this approach in describing some of the basic RHIC data \cite{Huovinen:2001cy,Teaney:2000cw,Teaney:2001av,Hirano:2002ds,Kolb:2002ve,Hama:2005dz,Eskola:2005ue,Nonaka:2006yn}, we use the recently developed hydrodynamic approach \cite{Chojnacki:2004ec,Chojnacki:2006tv,Chojnacki:2007jc} merged with the statistical-hadronization Monte-Carlo model \cite{Kisiel:2005hn} to globally describe the midrapidity hadron production at the highest RHIC energies and to make predictions for the future experiments at the Large Hadron Collider (LHC). Our approach uses standard methods of the field, however we take an effort to combine the best features for all entering elements of the existing analyses, except for the absence of elastic rescattering in the final state. We attempt a global fit to all main midrapidity observables, {\em i.e.}, the transverse-momentum spectra, the elliptic-flow coefficient, as well as the pionic HBT radii. The main outcome is that for pions and kaons a global fit works at the level of 10-15\%, which in our view is satisfactory, baring in mind systematic uncertainties in various elements of the approach, such as the determination of the initial condition, the lack of detailed knowledge of the equation of state, the assumed boost invariance, {\em etc.} The quality of the global fit is calling for predictions for the LHC energies, which is the main topic of this work. The predictions of other hydrodynamic models for phenomena expected at the LHC have been recently summarized in Ref.~\cite{Abreu:2007kv}.

Our implementation of inviscid hydrodynamics (Sect.~\ref{sect:eos}) incorporates the state-of-the-art knowledge of the hadronic equation of state. The relevant quantity is the sound velocity, $c_s$, considered as a function of temperature. The value of $c_s$ for the high-temperature strongly-interacting quark-gluon plasma is taken from the recent lattice simulations \cite{Aoki:2005vt}, which exhibit a substantial departure from the ideal-gas Stefan-Boltzmann limit even at temperatures significantly above the cross-over temperature \mbox{$T_c\sim 170$~MeV}. In particular, $c_s^2$ is significantly below the ideal-gas value of 1/3. At low temperatures the hadron-gas equation of state is used, with a complete treatment of resonances \cite{Broniowski:2000bj}. In the limit $T \to 0$ the function $c_s^2(T)$ approaches zero as $(T/m_\pi)^{1/2}$, characteristic of the massive pion gas at low temperatures.  In the cross-over region near $T_c$ a simple-minded interpolation between the high- and low-temperature equations of state is constructed. In accordance to the present knowledge, no phase transition but a {\em smooth cross-over} is built in, thus $c_s$ does not drop to zero at $T_c$. With the assumed features our hydrodynamics leads to smooth, laminar flow; no shock-waves are formed.     
  
Boost invariance is assumed for the considered {\em midrapidity} analysis, which is a good approximation for the highest RHIC and LHC energies at the rapidity window $|y|<1$ \cite{Bearden:2003fw,Bearden:2004yx}. This assumption reduces the number of independent space dimensions, which greatly simplifies the numerical analysis. The hydrodynamic equations are solved with the technique described in Refs.~\cite{Chojnacki:2004ec,Chojnacki:2006tv,Chojnacki:2007jc} which is a generalization of the method introduced by Baym et al. in Ref. \cite{Baym:1983sr}. The initial condition for the evolution is obtained from a Glauber profile (Sect.~\ref{sect:initcond}), with the initial central temperature $T_i$. The evolution proceeds until the temperature drops down to the freeze-out temperature $T_f$. These two temperatures are the free parameters of the approach. Entropy conservation is used as a numerical test, satisfied at the relative level of $10^{-4}$. The freeze-out hypersurface and the flow profile obtained from the hydrodynamic expansion are then used as input for the thermal event generator {\tt THERMINATOR} \cite{Kisiel:2005hn}. This code implements the statistical hadronization, accounting for a complete treatment of hadronic resonances (the included resonances and their branching ratios are the same  as in the {\tt SHARE} package \cite{Torrieri:2004zz}). Rescattering after the chemical freeze-out is not incorporated, which is an approximation working fairly well for the pions, as may be inferred from the results presented in Ref.~\cite{Nonaka:2006yn}. We note that a Monte Carlo generator of functionality similar to {\tt THERMINATOR} has been released recently \cite{Amelin:2006qe}.

Basic physical observables are calculated from the sample of events generated by {\tt THERMINATOR} (Sects.~\ref{sect:hadronization}, \ref{sec:RHIC} and \ref{sec:LHC}). For the highest RHIC energies the pion and kaon transverse-momentum spectra agree well with the data, with a similar quality matching of the elliptic flow coefficient $v_2(p_T)$. For the protons the model spectra are somewhat too steep and $v_2$ too large, which is probably due to the absence of the final-state elastic rescattering. The femtoscopic observables are evaluated  with the help of the two-particle method accounting for the effects of resonance decays and the Coulomb final-state interactions (Sects.~\ref{sect:corr}, \ref{sec:RHIC} and \ref{sec:LHC}). The use of the two-particle method imitating the experimental analysis is a clear advantage over the use of simple parameterizations of the emitting source \cite{Florkowski:2005nh}. The RHIC pionic HBT radii are reasonably reproduced in our approach, with $R_{\rm side}$ 10-15\% below the experimental data, $R_{\rm out}$ within the experimental errors, and $R_{\rm long}$ about 15\% too large. The ratio $R_{\rm out}/R_{\rm side}$ is about 1.2-1.25 and almost constant as a function of the the transverse momentum of the pair, which is still away from the experimental ratio, but considerably better than in many other hydrodynamic approaches. All in all, at the highest RHIC energies the approach gives in our view a quite satisfactory global description of the soft pionic observables, which agree with the data at the level of 10-15\%. Some discrepancies may be attributed to 
the final-state elastic interactions among hadrons, not included in our description. 

Extrapolating $T_i$ to higher energies allows us to make global prediction for the soft hadronic physics at the LHC, which are presented in Sect.~\ref{sec:LHC}. We start the hydrodynamic evolution from a higher initial temperatures, using a few values in the expected range: $T_i=400$, $450$, and $500$~MeV, and keep all other parameters fixed, in particular $T_f=150$~MeV. We observe the expected growth of particle multiplicities and the increase of the flow, resulting in smaller slope of the $p_T$-spectra. The elliptic flow of pions does not increase any more, showing the expected saturation of $v_2$, and even decreases slightly at the highest tested temperature of $T_i=500$~MeV. The HBT radii increase moderately with the temperature: $R_{\rm side}$ and $R_{\rm long}$ roughly 1~fm for each 100~MeV increase of $T_i$, and $R_{\rm out}$ even slower. We provide some simple formulas describing the change of the considered quantities with $T_i$.

Finally, we remark that {\tt THERMINATOR} may be used straightforwardly to test the soft physics in detector simulations at LHC. It produces particles with full decay history which can be fed directly to transport codes. It includes non-trivial and physically well-motivated predictions for flow phenomena (both the radial and elliptic flow) with particle type dependence naturally built in. It also provides  information on emission points for femtoscopic simulation. The non-trivial, dynamic (containing space-momentum correlations) emission function allows for the testing more advances femtoscopic analysis techniques,  {\em e.g.} spherical harmonics decompositions or imaging 
\cite{Brown:1997ku,Danielewicz:2007jn,Vertesi:2007ki,Brown:2007raa}.

\begin{figure}[tb]
\begin{center}
\includegraphics[angle=0,width=0.4 \textwidth]{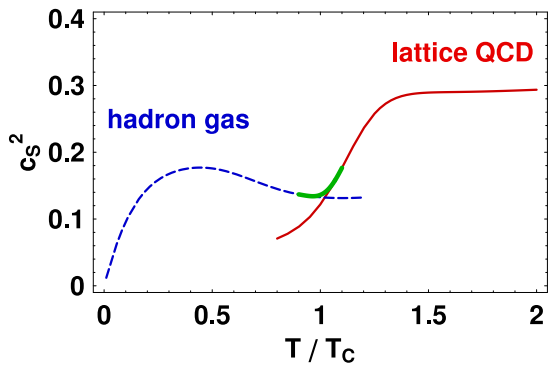}
\end{center}
\vspace{-3mm}
\caption{(Color online) The square of the sound velocity as a function of temperature. The high-temperature part comes from the lattice simulations of Ref.~\cite{Aoki:2005vt}, the low-temperature part from the hadronic gas with resonances, while the transition region near the cross-over is obtained with simple interpolation. Here we take $T_c = 170$~MeV as the scale for $T$.}
\label{fig:cs2}
\end{figure}

\section{Equation of state and perfect-fluid hydrodynamics}
\label{sect:eos}

\begin{figure}[tb]
\begin{center}
\includegraphics[angle=0,width=0.475 \textwidth]{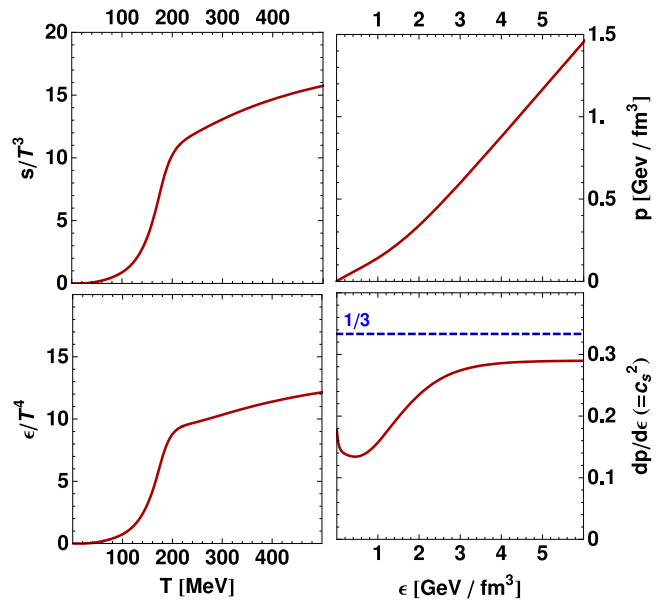}
\end{center}
\vspace{-3mm}
\caption{(Color online) The two left panels: the volume densities of entropy and energy, scaled by $T^3$ and $T^4$, respectively, 
shown as functions of the temperature. The two right panels: the pressure and sound velocity shown as functions of the energy density. 
The presented thermodynamic functions follow directly from the temperature-dependent sound velocity shown in Fig.~\ref{fig:cs2}. }
\label{fig:thermo}
\end{figure}

In our studies we focus on the description of the midrapidity region. Statistical models applied to the highest-energy RHIC data yield the baryon chemical potential $\mu_B \approx 20-30~{\rm MeV}$ \mbox{\cite{Florkowski:2001fp,Braun-Munzinger:2001ip,Baran:2003nm,Cleymans:2004pp,Biedron:2006vf}}, while the predictions for LHC energies  give \mbox{$\mu_B \approx$ 0.8 MeV} \cite{Andronic:2005yp}. Hence, for the hydrodynamic equations we can approximately assume that the baryon chemical potential vanishes. In this situation the whole information on the equation of state is encoded in the temperature-dependent sound velocity $c_s(T)$ \cite{Chojnacki:2004ec}. In the hydrodynamic calculations we use the function $c_s(T)$ introduced in Ref.~\cite{Chojnacki:2007jc} and labeled there as ``case~I''. At low temperatures it is given by the sound velocity of a hadron gas (with a complete set of hadronic resonances). The function $c_s^2(T)$ approaches zero as $(T/m_\pi)^{1/2}$, characteristic of the pion gas. At high temperatures the equation of state coincides with the recent lattice simulations \cite{Aoki:2005vt}. In the intermediate-temperature region we take the simple smooth interpolation between the hadron gas and lattice results, see Fig.~\ref{fig:cs2}. As mentioned above, the knowledge of the function $c_s(T)$ allows us to determine all other thermodynamic properties of our system. In Fig.~\ref{fig:thermo} we display the corresponding entropy and energy densities as functions of $T$, as well as the pressure and the sound velocity as functions of the energy density. The region in the energy density where $c_s^2$ drops to zero is exceedingly small and invisible in Fig.~\ref{fig:thermo}, while the full dependence can be seen in Fig.~\ref{fig:cs2}.   
Other equations of state and their impact on the physical observables were recently studied in Ref.~\cite{Huovinen:2005gy}.

We stress that by using the lattice results we take into account the non-perturbative aspects of the plasma behavior, which may be regarded as the effective inclusion of the strongly-interacting quark-gluon plasma; large deviations from the ideal-gas behavior directly indicate the non-negligible interactions present in the plasma. In particular, $c_s^2$ is significantly below the ideal-gas value of 1/3 also at temperatures way above $T_c$. Note that in agreement with the present knowledge, no phase transition is present in the system, but a {\em smooth cross-over}, therefore $c_s$ does not drop to zero at $T_c$ but remains a smooth function. Also, the criterion for absence of shock-waves is satisfied \cite{Baym:1983sr,Blaizot:1987cc}, which makes the evolution simple  to solve numerically -- we use the adaptive method of lines in the way implemented in the MATHEMATICA package. The space directions are discretized, and the integration in time is treated as solving of the system of ordinary differential equations with the adaptive time step. Entropy conservation is used as a numerical test, 
which is satisfied at the relative level of $10^{-4}$. 

With the assumed features our hydrodynamics leads to smooth, laminar flow.  In Ref. \cite{Chojnacki:2007jc} it was argued that the use of the sound-velocity function depicted in Fig. \ref{fig:cs2} is physically attractive, since it leads to relatively short time scales of the hydrodynamic evolution. This, in turn, helps to obtain the satisfactory description of the HBT radii at the highest RHIC energies.

\begin{figure}[tb]
\begin{center}
\includegraphics[angle=0,width=0.24 \textwidth]{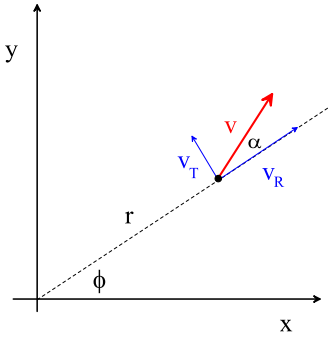}
\end{center}
\vspace{-3mm}
\caption{(Color online) The notation for the position and velocity of the fluid element in the transverse plane.}
\label{fig:alpha}
\end{figure}

The relativistic hydrodynamic equations of the perfect fluid follow from the energy-momentum conservation and the assumption of local equilibrium. For the baryon-free matter they read
\begin{eqnarray}
u^\mu \partial_\mu (T u^\nu) &=& \partial^\nu T, 
\label{h1} \\
\partial_\mu (s u^\mu) &=& 0, 
\label{h2}
\end{eqnarray}
where $T$ is the temperature, $s$ the entropy density, and $u^\mu = \gamma(1,{\bf v})$ denotes the four-velocity of the fluid. The normalization condition $u_\mu u^\mu=1$ implies that only three out of four equations appearing in (\ref{h1}) are independent. As mentioned above, in the analysis of the evolution in the midrapidity region we additionally assume that the system is boost-invariant. This assumption introduces another constraint, hence, in this case Eqs.~(\ref{h1}) and (\ref{h2}) reduce to three independent equations, which may be written in the following form \cite{Dyrek:1984xz}
\begin{eqnarray}
\frac{\partial }{\partial t}\left( rts\gamma \right) +\frac{\partial }{\partial r}
\left( rts\gamma v\cos \alpha \right) + \frac{\partial }{\partial
\phi }\left( ts\gamma v\sin \alpha \right)  &=&0, 
\nonumber \\
\frac{\partial }{\partial t}\left( rT\gamma v\right) +
r\cos \alpha \frac{\partial }{\partial r}\left( T\gamma \right) 
+\sin \alpha \frac{\partial }{\partial \phi }\left( T\gamma \right)  &=&0, 
\nonumber \\
T\gamma ^{2}v\left( \frac{d\alpha }{dt}+\frac{v\sin \alpha }{r}\right) -\sin
\alpha \frac{\partial T}{\partial r}
+\frac{\cos \alpha }{r}\frac{\partial T}{\partial \phi } &=&0. \nonumber \\
\label{wfdyr}
\end{eqnarray}
Here $t, \, r\!\!=\!\!\sqrt{x^2+y^2}$, and $\phi=\hbox{tan}^{-1} (y/x)$ are the time and space coordinates which parameterize the 
transverse plane $z=0$ (for boost-invariant systems, the values of all physical quantities at $z \neq 0$ may be obtained by the 
Lorentz transformation). The quantity $v$ is the transverse flow, $\gamma =1/\sqrt{1-v^2}$ is the Lorentz factor, and $\alpha$ is the 
dynamically determined angle between the direction of the transverse flow and the radial direction, see Fig.~\ref{fig:alpha}. The 
differential operator $d/dt$ represents the complete time derivative and is defined by the formula
\begin{equation}
\frac{d}{dt} = \frac{\partial}{\partial t} 
+  v \cos\alpha \frac{\partial}{\partial r} 
+ \frac{v \sin\alpha}{r} \frac{\partial}{\partial \phi}.
\end{equation} 
Equations (\ref{wfdyr}) are three equations for four unknown functions: $T$, $s$, $v$, and $\alpha$. To obtain a closed system of equations one
needs to supplement them with an equation of state, {\em i.e.}, with a relation connecting $T$ and $s$. Alternatively, one may fix the
temperature-dependent sound velocity,
\begin{equation}
c_s^2(T) = \frac{\partial P}{\partial \epsilon} = 
\frac{s}{T}\frac{\partial T}{\partial s},
\label{cs}
\end{equation}
to a known function. This method is advantageous, as it directly uses the available information on the sound velocity.

\section{Initial conditions}
\label{sect:initcond}

Similarly to other hydrodynamic calculations, we assume that the initial entropy density of the particles produced at the transverse position point $\vec{x}_{T }$ is proportional to a profile obtained from the Glauber approach. Specifically, we use a mixed model \cite{Back:2001xy,Back:2004dy}, with a linear combination of the wounded-nucleon density $\rho_{\rm W}\left( \vec{x}_{T } \right)$ and the density of binary collisions $\rho_{\rm bin}\left( \vec{x}_{T } \right)$, namely 
\begin{equation}
s \left( \vec{x}_{T } \right)   \propto 
\rho \left( \vec{x}_{T } \right) = 
\frac{1-\kappa}{2} \, \rho_{\rm W}\left( \vec{x}_{T } \right) + \kappa \rho_{\rm bin} \left( \vec{x}_{T } \right). 
\label{initialeps2}
\end{equation}
The case $\kappa=0$ corresponds to the standard wounded-nucleon model \cite{Bialas:1976ed}, while $\kappa=1$ would include  the binary collisions only. The PHOBOS analysis \cite{Back:2004dy} of the particle multiplicities yields $\kappa=0.12$ at $\sqrt{s_{NN}}=17~{\rm GeV}$ and $\kappa=0.14$ at $\sqrt{s_{NN}}=17~{\rm GeV}$. In this paper we assume 
\begin{eqnarray}
\kappa=0.14 \label{eq:A}
\end{eqnarray}
for both the highest RHIC and the LHC energies. Since the density profile from the binary collisions 
is steeper than from the wounded nucleons, increased values of $\kappa$ yield steeper density profiles, which in turn result in steeper temperature profiles.

In the hydrodynamic code, the initial conditions are specified for the temperature profile which, according to Eq.~(\ref{initialeps2}), 
takes the form
\begin{equation}
T(\tau_{\rm init},\vec{x}_{T } ) = T_S \left[ s_i
\frac{ \rho \left( \vec{x}_{T } \right) }{\rho \left(0 \right)} \right],
\label{Tt0}
\end{equation}
where $T_S(s)$ is the inverse function to the function $s(T)$, and $s_i$ is the initial entropy at the center of the system. The initial central temperature $T_i$ equals $T_S(s_i)$. Throughout this paper we take the initial time for the hydrodynamic evolution to be
\begin{eqnarray}
\tau_{\rm init} = 1~{\rm fm}. \label{tauinit}
\end{eqnarray}

The wounded-nucleon and the binary-collisions densities in Eq.~(\ref{Tt0}) are obtained from the optical limit of the Glauber model, which is a very good approximation for not too peripheral collisions \cite{Miller:2007ri}. The standard formulas are \cite{Bialas:1976ed}
\begin{eqnarray}
& &\!\!\!\!\!\!\!\! \rho_{\rm W} \left( \vec{x}_{T } \right) = \nonumber \\
& & 
T_A\left(\frac{ \vec{b} }{2}+ { \vec{x}_{T } } \right)
\left\{1\! -\! \left[1\! -\! \frac{\sigma}{A} \, 
T_A\left(-\frac{\vec{b} }{2}+ {\vec{x}_{T } } \right)
\right]^A \right\} \nonumber \\
& &\!\!\!\!\!\!\!\! + \, T_A\left(-\frac{\vec{b} }{2}+ {\vec{x}_{T } } \right)
\left\{1\! -\! \left[1\! -\! \frac{\sigma}{A} \, 
T_A\left(\frac{\vec{b} }{2}+ {\vec{x}_{T } } \right)
\right]^A \right\} \nonumber \\
\label{rhoWN}
\end{eqnarray}
and
\begin{equation}
\rho_{\rm bin} \left( \vec{x}_{T } \right) = \sigma
T_A\left(\frac{ \vec{b} }{2}+ { \vec{x}_{T } } \right)
T_A\left( - \frac{ \vec{b} }{2}+ { \vec{x}_{T } } \right).
\label{rhoBC}
\end{equation}
In Eqs. (\ref{rhoWN}) and  (\ref{rhoBC}) $\vec{b}$ is the impact parameter, $\sigma$ is the nucleon-nucleon total inelastic cross section, and $T_A\left(x,y\right)$ denotes the nucleus thickness function 
\begin{equation}
T_A(x,y) = \int dz \, \rho\left(x,y,z\right).
\label{TA}
\end{equation}
For RHIC energies we use the value $\sigma = 42~{\rm mb}$, while for LHC we take
$\sigma = 66~{\rm mb}$. The function $\rho(r)$ in Eq. (\ref{TA}) is the nuclear density profile given by the Woods-Saxon function with the conventional choice of  
parameters: 
\begin{eqnarray}
\rho_0 &=& 0.17 \,\hbox{fm} ^{-3},  \nonumber \\
r_0 &=& (1.12 A^{1/3} -0.86 A^{-1/3}) \,\hbox{fm},  \nonumber \\
a &=& 0.54 \,\hbox{fm}.
\label{woodssaxon}
\end{eqnarray}
The atomic number $A$ is 197 for RHIC (gold nuclei) and 208 for LHC (lead nuclei). The value of the impact parameter in Eqs. (\ref{rhoWN}) and  (\ref{rhoBC}) depends on the considered centrality class. 

We stress that the shape of the initial condition (\ref{Tt0}) is important, as it determines the development of the radial and elliptic flow, thus affecting such observables as the $p_T$-spectra, $v_2$, and the femtoscopic features. On qualitative grounds, sharper profiles lead to more rapid expansion. Several effects should be considered here. Firstly, as discussed in Ref.~\cite{Kolb:2000sd}, hydrodynamics starts a bit later, when the profile is less eccentric than originally due to early evolution of partons in the pre-hydro phase. On the other hand, statistical fluctuations of the axes of the second harmonic in the distribution of Glauber sources (wounded nucleons, binary collisions)
\cite{Aguiar:2000hw,Aguiar:2001ac,Socolowski:2004hw,Voloshin:2006gz,Broniowski:2007ft,Broniowski:2007nz,Voloshin:2007pc,Alver:2007cs} leads to a significant enhancement of the eccentricity, especially at low values of the impact parameter. Thus the initial eccentricity may in fact be smaller or larger than what follows from the application of the Glauber model. This contributes to the systematic model uncertainty at the level of, say, 10-20\%. This uncertainty could only be reduced by the employment of a realistic model of the pre-hydrodynamic evolution and is outside of the scope of this work. With this uncertainty in place, one should not expect or demand a better agreement with the physical observables than at the corresponding level of 10-20\%.

\begin{figure}[t]
\begin{center}
\includegraphics[angle=0,width=0.25 \textwidth]{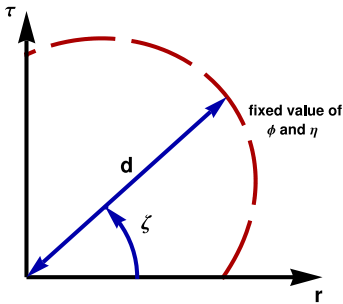}
\end{center}
\vspace{-3mm}
\caption{(Color online) The parameterization of the freeze-out hypersurface.}
\label{fig:zeta}
\end{figure}

\section{Statistical hadronization with THERMINATOR}
\label{sect:hadronization}

In order to calculate the physical observables we first use the hydrodynamic code with the described initial conditions to determine the freeze-out hypersurface $\Sigma$. In our approach it is defined by the condition of a constant 
freeze-out temperature $T_f$, and parameterized in the general way by the following equations
\begin{eqnarray}
t &=& d \left( \phi ,\zeta, \eta \right) \sin \zeta \,\cosh \eta ,\quad
z= d\left( \phi ,\zeta, \eta \right) \sin \zeta \, \sinh \eta , \nonumber \\
x &=& d\left( \phi ,\zeta, \eta\right) \cos \zeta \, \cos \phi ,\quad
y = d\left( \phi ,\zeta, \eta \right) \cos \zeta \, \sin \phi. \nonumber \\
\label{cooperfryeparam}
\end{eqnarray}
The variable $\eta$ is the space-time rapidity defined as usual by the formula
\begin{equation}
\eta = \frac{1}{2} \ln \frac{t+z}{t-z} = \tanh^{-1} \left(\frac{z}{t}\right).
\end{equation}
The parameterization (\ref{cooperfryeparam}) yields the compact expressions for the proper time $\tau$ and the transverse distance $r$, 
\begin{eqnarray}
\tau &=& \sqrt{t^2 - z^2} = \,d\left( \phi ,\zeta, \eta \right) \sin \zeta , \nonumber \\
r &=& \sqrt{x^2+y^2}= \,d\left( \phi ,\zeta, \eta \right) \cos \zeta  .
\label{tauandrho}
\end{eqnarray}
At any given value of the space-time rapidity $\eta$ the position of the point on the hypersurface $\Sigma$ is defined by the two angles, $\phi $ and $\zeta$, and the distance from the origin of the coordinate system, $d\left(\phi ,\zeta, \eta \right)$. The angle $\phi$ is the standard azimuthal angle in the $y-x$ plane as used in the hydrodynamic equations (\ref{wfdyr}), while the angle $\zeta$ is the azimuthal angle in the $\tau-r$ plane. We have introduced the angle $\zeta$ because in most cases the freeze-out curves in the $\tau-r$ plane  may be treated as functions of this parameter. The use of the transverse distance $r$ is inconvenient, since very often two freeze-out points correspond to one value of $r$, see Fig. \ref{fig:zeta}.

With the standard definition of the four-momentum in terms of the rapidity and transverse momentum, 
\begin{equation}
p^\mu = \left( m_T \cosh y, p_T \cos\phi_p, p_T \sin\phi_p, m_T \sinh y \right).
\end{equation}
where $m_T = \sqrt{m^2+p_T^2}$ is the transverse mass, and with the standard definition of the element of the hypersurface $d\Sigma _{\mu }$, we find the explicit form of the Cooper-Frye integration measure \cite{Cooper:1974mv}
\begin{widetext}
\begin{eqnarray}
&&d\Sigma _{\mu }\,p^{\mu } =  d^{\,2}\sin \zeta \left[ 
\vphantom{\frac{\partial d}{\partial \eta } }
d\cos \zeta
\,\left( m_{T }\sin \zeta \cosh \left( \eta -y\right) +p_{T }\cos
\zeta \cos \left( \phi -\phi _{p}\right) \right) \right.  \label{dSigma1} \\
&& +\frac{\partial d}{\partial \zeta }\cos \zeta \,\left( -m_{T
}\cos \zeta \cosh \left( \eta -y\right) +p_{T }\sin \zeta \cos
\left( \phi -\phi _{p}\right) \right) + \frac{\partial d}{\partial \phi }  p_{T }\sin \left( \phi -\phi
_{p}\right) \left. + \frac{\partial d}{\partial \eta } \, \cot \zeta \,
 m_T \sinh \left( \eta - y\right)
\right] d\eta d\phi d\zeta .\nonumber
\end{eqnarray}
This equation, when used in the Cooper-Frye formula \cite{Cooper:1974mv}, leads to the six-dimensional particle density at freeze-out
\begin{eqnarray}
\frac{dN}{dy d\phi_p p_T dp_T d\eta d\phi d\zeta} & = & g \frac{d^{\,2}\sin \zeta}{(2\pi)^3} \left[ 
\vphantom{\frac{\partial d}{\partial \eta } }
d\cos \zeta
\,\left( m_{T }\sin \zeta \cosh \left( \eta -y\right) +p_{T }\cos
\zeta \cos \left( \phi -\phi _{p}\right) \right) \right.  \nonumber \\
& & + \left. \frac{\partial d}{\partial \zeta }\cos \zeta \,\left( -m_{T
}\cos \zeta \cosh \left( \eta -y\right) +p_{T }\sin \zeta \cos
\left( \phi -\phi _{p}\right) \right) + \frac{\partial d}{\partial \phi } \,
p_{T }\sin \left( \phi -\phi _{p}\right) \right]  \nonumber \\
& & \times \left\{\exp \left[ 
\frac{\beta m_T}{\sqrt{1-v^2}} \cosh(y-\eta)
-\frac{\beta p_T v}{\sqrt{1-v^2}} \cos(\phi+\alpha-\phi_p) - \beta \mu \right] \pm 1 \right\}^{-1}.
\label{6cf1}
\end{eqnarray}
\end{widetext}
In the transition from (\ref{dSigma1}) to (\ref{6cf1}) we made use of the boost-invariance and removed the term $\partial d/ \partial \eta$, since the function $d$ depends only on $\phi$ and $\zeta$. The last line in (\ref{6cf1}) contains the equilibrium statistical distribution function, with  $-1$ for bosons and $+1$ for fermions. The argument is the Lorentz-invariant product $p^\mu u_\mu$. The parameter \mbox{$\beta = 1/T$} is the inverse temperature and $g$ denotes the spin degeneracy factor. The four-velocity field has been expressed in terms of the transverse flow $v$ and the dynamical angle $\alpha$ which depend on the space-time positions on the freeze-out hypersurface. The statistical distribution functions include the chemical potential, 
\begin{eqnarray}
\mu =\mu_B B + \mu_S S+ \mu_{I_3} I_3, \label{mus}
\end{eqnarray}
with $B$, $S$, and $I_3$ denoting the baryon number, strangeness, and isospin of the particle, respectively, while the baryonic, strange, and isospin chemical potentials assume the values: $\mu_B=28.5~{\rm MeV}$, $\mu_S=9~{\rm MeV}$, $\mu_{I_3}=-0.9~{\rm MeV}$ at the highest RHIC energies \cite{ Baran:2003nm}, and $\mu_B= 0.8~{\rm MeV}$ and $\mu_S=\mu_{I_3}=0~{\rm MeV}$ at the LHC \cite{Andronic:2005yp}.

In order to describe the statistical hadronization and production of particles we first run our hydrodynamic code and extract the functions describing the freeze-out hypersurface and flow: $d=d(\phi,\zeta)$, $v=v(\phi,\zeta)$, and $\alpha=\alpha(\phi,\zeta)$. In the next step these functions are used as input for the {\tt THERMINATOR} code, which generates, according to the formula (\ref{6cf1}), the distributions of the primordial particles. The primordial particles include the stable hadrons as well as all hadronic resonances. The resonances decay through strong, electromagnetic, or weak interactions at the random proper time controlled by the particle's life-time, and at the location following from the kinematics. As the final result we obtain the stable particle distributions including 
the feeding from the resonance decays. Since the memory on all decays is kept, one may apply the experimental cuts or the weak-decay feeding policy, which facilitates the more accurate comparison of the model to the data. For example, one may extract easily the model proton spectrum which does not include the feeding from the $\Lambda$ decays. 

The momentum distribution functions are obtained by integrating over $\phi$, $\zeta$, and $\eta$ 
\begin{equation}
\frac{dN}{dy d^2 p_T} = \int\limits_{-\infty}^{\infty} d\eta
\int\limits_{0}^{\pi} d\zeta  \int\limits_{0}^{2 \pi} d\phi
\frac{dN}{dy d\phi_p p_T dp_T d\eta d\phi d\zeta}.
\label{cf1}
\end{equation}
For cylindrically asymmetric collisions and midrapidity, $y=0$, the transverse-momentum spectrum has the following expansion in the azimuthal angle of the emitted particles
\begin{equation}
\frac{dN}{dy d^2 p_T} = \frac{dN}{dy \, 2 \pi p_T \, d p_T }
\left( 1 + 2 v_2(p_T)  \cos(2 \phi_p) + ...\right).
\label{v2def}
\end{equation}
Eq. (\ref{v2def}) defines the elliptic flow coefficient $v_2$. 

The well-known problem of the freeze-out condition in the Cooper-Frye formulation is that the hypersurface from hydrodynamics  typically contains a non-causal piece, where particles are emitted 
back to the hydrodynamic region \cite{Bugaev:2004kq}. In our approach we follow the usual strategy of including only that part of the hypersurface where $d\Sigma_\mu p^\mu \ge 0$. To estimate the effect from the non-causal part we compute the ratio of particles flowing backwards ({\em i.e.} where $d\Sigma_\mu p^\mu < 0$) to all particles. For the cases studied in the following sections we find this ratio to be a fraction of a percent. Such very small values show that the known conceptual problem is not of practical importance for our study.

\section{Pion correlation function}
\label{sect:corr}

The  correlation function for identical pions is obtained with the two-particle Monte-Carlo method discussed in 
detail in Ref.~\cite{Kisiel:2006is,Kisiel:2006yv}. In this approach the evaluation of the correlation function is reduced to the 
calculation of the following expression
\begin{eqnarray}
&& \!\!\!\!\!\!C({\vec q}, {\vec k}) = \nonumber \\
&& \!\!\!\!\!\!\frac{\sum\limits_{i} \sum\limits_{j \neq i} \delta_\Delta({\vec q} 
- {\vec p}_i + {\vec p}_j ) \delta_\Delta({\vec k} - \frac{1}{2}({\vec p}_{i} + {\vec p}_{j}) )
|\Psi({\vec k}^{*}, {\vec r}^{*}) |^2} 
{\sum\limits_i \sum\limits_{j \neq i} \delta_\Delta({\vec  q} - {\vec p}_i + {\vec p}_j ) 
\delta_\Delta({k} - \frac{1}{2}({\vec p}_{i} + {\vec p}_{j} ))}, \nonumber \\
\label{cfbysum}
\end{eqnarray}
where $\delta_{\Delta}$ denotes the box function
\begin{eqnarray}
\delta_{\Delta}({\vec p}) = 
\left\{
\begin{array}{cc}
1 & \hbox{if}  \,\,\,  | {p_{x}} | \leq  \frac{\Delta}{2} ,| {p_{y}} | \leq  \frac{\Delta}{2},  | {p_{z}} | \leq  \frac{\Delta}{2} \\
& \\
0 & \hbox{otherwise}.
\end{array}
\right.
\label{deltadelta}
\end{eqnarray}     
In the numerator of Eq. (\ref{cfbysum}) we include the sum of the squares of modules of the wave function 
calculated for all pion pairs with the relative momentum $\vec q$ (we use the bin resolution $\Delta = 5~{\rm MeV})$ and the pair average momentum $\vec k$. For non-central collisions we only provide azimuthally integrated HBT radii. For each pair the wave function $\Psi({\vec k}^{*}, {\vec r}^{*})$, including the Coulomb interaction, is calculated in the rest frame of the pair; ${\vec k}^{*}$ and ${\vec r}^{*}$ denote the relative momentum and the relative distance in the pair rest frame, respectively. In the denominator of Eq. (\ref{cfbysum}) we put the number of pairs with the relative momentum $\vec q$ and the average momentum $\vec k$. The correlation function (\ref{cfbysum}) is then expressed with the help of the Bertsch-Pratt coordinates $k_T, q_{\rm out}, q_{\rm side}, q_{\rm long}$ and approximated by the Bowler-Sinyukov formula \cite{Bowler:1991vx,Sinyukov:1998fc}
\begin{eqnarray}
&& C(\vec q, \vec k) = (1 - \lambda) + \lambda K_{\rm coul}(q_{\rm inv})
\left[1 +  \exp \left(-R_{\rm out}^2 q_{\rm out}^2  \right. \right.
\nonumber \\
&& \left. \left.
- R_{\rm side}^2 q_{\rm side}^2 - R_{\rm long}^2
q_{\rm long}^2 \right)  \right],
\label{cffitbs}
\end{eqnarray}
where $K_{\rm coul}(q_{\rm inv})$ with $q_{\rm inv} = 2 k^*$ is the squared Coulomb wave function integrated over a static gaussian source. 
We use, following the STAR procedure \cite{Adams:2004yc}, the static gaussian source characterized by the widths of 5 fm in all three directions. 
Four $k_{T}$ bins, $(0.15-0.25)$, $(0.25-0.35)$, $(0.35-0.45)$, and $(0.45-0.60)$~GeV, are considered.
The 3-dimensional correlation function with the exact treatment of the Coulomb interaction is then fitted with this approximate formula and the HBT radii $R_{\rm out}$, $R_{\rm side}$, and $R_{\rm long}$ are obtained. They can be compared directly to the experimental radii.


\begin{figure}[tb]
\begin{center}
\includegraphics[angle=0,width=0.4 \textwidth]{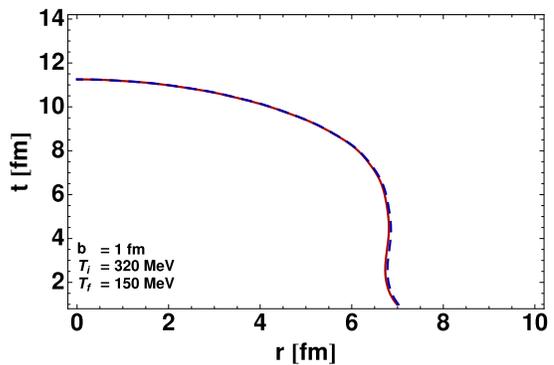}
\end{center}
\caption{(Color online) In-plane and out-of-plane freeze-out curves, {\em i.e.}, 
the intersections of the freeze-out hypersurface with the planes $y=0$ and $x=0$, obtained for central RHIC collisions; 
$b = 1~{\rm fm}$,  $T_i = 320~{\rm MeV}$, and $T_f = 150~{\rm MeV}$. The two curves overlap, 
indicating that the system at freeze-out is almost azimuthally symmetric in the transverse plane. }
\label{fig:rhic-cent-HS}
\end{figure}

\begin{figure}[tb]
\begin{center}
\includegraphics[angle=0,width=0.45 \textwidth]{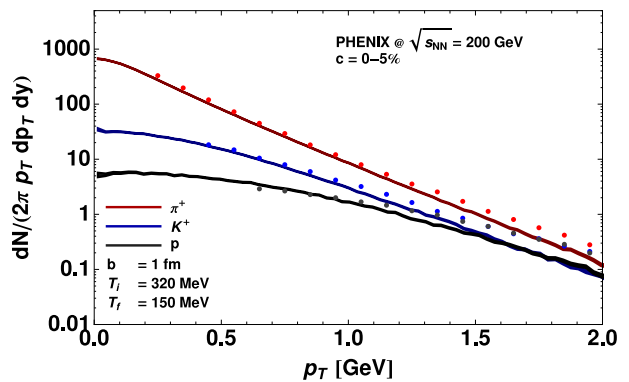}
\end{center}
\caption{(Color online) Transverse-momentum spectra of $\pi^+$, $K^+$, and protons. 
The PHENIX experimental results \cite{Adler:2003cb} for Au+Au collisions at $\sqrt{s_{NN}}= 200~{\rm GeV}$ and the centrality class 0-5\% (points) are compared to the model calculations (solid lines) with the same parameters as in Fig.~\ref{fig:rhic-cent-HS}. }
\label{fig:rhic-cent-pTsp}
\end{figure}

\begin{figure}[tb]
\begin{center}
\includegraphics[angle=0,width=0.4\textwidth]{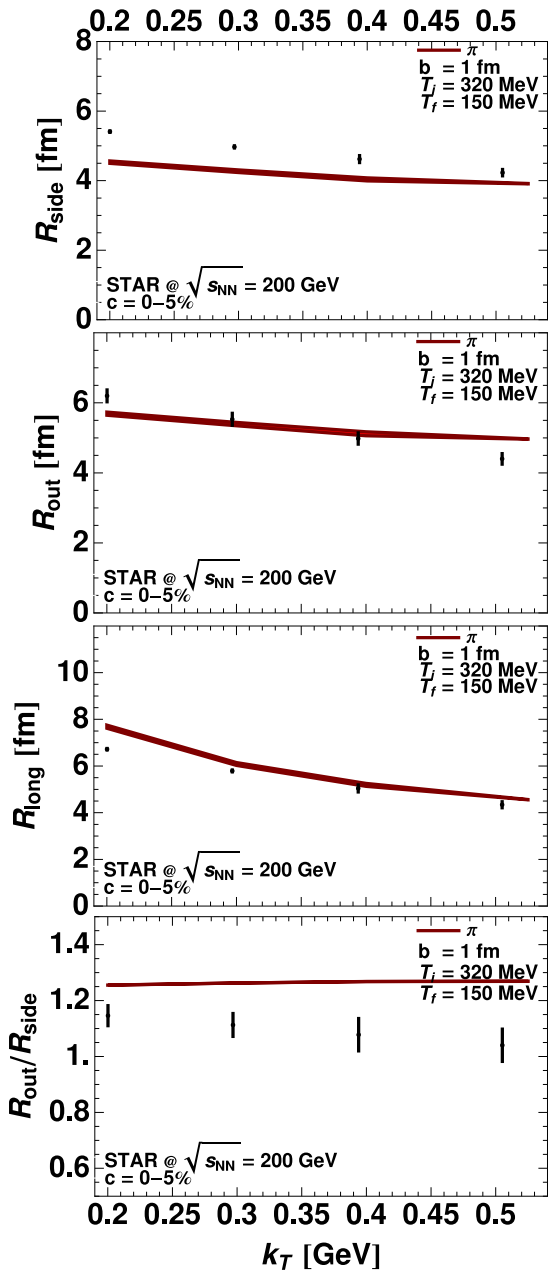}
\end{center}
\caption{The pionic HBT radii plotted as functions of the average transverse momentum of the pair compared to the STAR data \cite{Adams:2004yc} at the centrality 0-5\%. The calculation uses the two-particle method and includes the Coulomb effects.  The values of the model parameters are the same as in Fig.~\ref{fig:rhic-cent-HS}.}
\label{fig:rhic-cent-HBT}
\end{figure}

\section{Description of the highest-energy RHIC data}
\label{sec:RHIC}

In this Section we apply our model to describe the soft hadron production in the relativistic heavy-ion collisions studied at the highest RHIC energy, \mbox{$\sqrt{s_{NN}}$ = 200 GeV}, by the PHENIX \cite{Adler:2003cb,Adler:2003kt} and STAR Collaborations \cite{Adams:2004yc}. We consider 
the most central events, given by the centrality class 
\mbox{$c$ = 0 - 5\%}, where we use the impact parameters $b=1$~fm.
We also use the data from the centrality classes 
\mbox{$c$ = 20 - 30\%} (transverse-momentum spectra and the HBT radii) and \mbox{$c$ = 20 - 40\%} (elliptic flow), for which we take 
(for simplicity) the same value of the impact parameter, $b=7$~fm \footnote{In a detailed analysis a more accurate study in each centrality 
class should be made. Note, however, that the link between the centrality class and the impact parameter involves the 
total inelastic cross section, and to a very good accuracy for not-too-peripheral collisions 
$c \simeq \pi b^2/\sigma_{\rm inel.}$ \cite{Broniowski:2001ei}. The total inelastic cross section is not measured, instead is 
obtained from Glauber simulations, thus carries model uncertainty \cite{Miller:2007ri}.}.

\subsection{Central collisions}
\label{sec:RHIC-central}

First, we consider the centrality class 0 - 5\% with the corresponding value of the impact parameter \mbox{$b = 1$~fm}. The in-plane and out-of-plane freeze-out curves are defined as the intersections of the freeze-out hypersurface with the planes $y=0$ and $x=0$. They are obtained with the initial central temperature $T_i = 320~{\rm MeV}$ and the final (freeze-out) temperature $T_f = 150~{\rm MeV}$ and are shown in Fig.~\ref{fig:rhic-cent-HS}. The two freeze-out curves practically overlap, indicating that the expansion of the system is almost azimuthally symmetric in the transverse plane.  This effect is certainly expected for the almost central collisions, where the impact parameter is very small. 
We note that the shape of the isotherms is consistent with the result presented in Fig.~2 of Ref.~\cite{Eskola:2005ue}, where the same decoupling temperature of 150~MeV was considered.

In Fig.~\ref{fig:rhic-cent-pTsp} we present our results for the hadron transverse-momentum spectra with the same values of the parameters. The dots show the PHENIX data \cite{Adler:2003cb} for positive pions, positive kaons, and protons, while the solid lines show the results obtained from our hydrodynamic code connected to {\tt THERMINATOR}. Our model describes the data properly up to the transverse-momentum values of about 1.5~GeV. For larger values of $p_T$ the model underpredicts the data. This effect may be explained by the presence of the semi-hard processes, not included in our approach.

In our model, the values of $T_i$ and $b$ control the overall normalization. On the other hand, the value of the freeze-out temperature $T_f$ determines inter alia the relative normalization and the slopes of the spectra. We note that the correct slope for the pions and kaons is recovered at a relatively high value of $T_f$. This is possible, since the spectra of the observed hadrons contain the contributions from all well established hadronic resonances. This single-freeze-out picture \cite{Florkowski:2001fp,Broniowski:2001we,Broniowski:2001uk}, where the same temperature is used to describe the ratios of hadronic abundances and the spectra, was first tested for hadronic spectra in Ref.~\cite{Broniowski:2001we}, see also \cite{Rafelski:2000by,Broniowski:2002nf}. 
We note that the value \mbox{$T_f$ = 150 MeV} agrees with the recent results obtained in the framework of the single-freeze-out model in Ref.~\cite{Prorok:2007xp}. In the calculation of the proton spectra, in order to be consistent with the PHENIX experimental procedure, we removed the protons coming from the $\Lambda$ decays. 


\begin{figure}[tb]
\begin{center}
\includegraphics[angle=0,width=0.4 \textwidth]{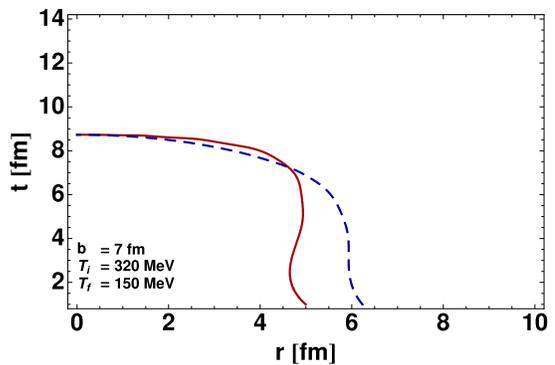}
\end{center}
\caption{(Color online) The freeze-out curves for \mbox{$T_i = 320~{\rm MeV}$}, \mbox{$T_f = 150~{\rm MeV}$}, and \mbox{$b = 7~{\rm fm}$}. The solid line describes the in-plane profile, while the dashed line describes the out-of-plane profile.}
\label{fig:rhic-pery-HS}
\end{figure}

\begin{figure}[tb]
\begin{center}
\includegraphics[angle=0,width=0.45 \textwidth]{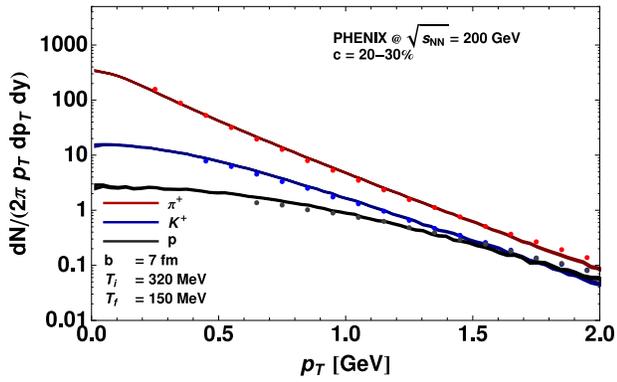}
\end{center}
\caption{(Color online) Transverse-momentum spectra of $\pi^+$, $K^+$, and protons. 
The PHENIX experimental results \cite{Adler:2003cb} for Au+Au collisions at $\sqrt{s_{NN}}= 200~{\rm GeV}$ and the centrality class 20-30\% (points) 
are compared to the model calculations (solid lines). The values of the model parameters are the same as in Fig. \ref{fig:rhic-pery-HS}. }
\label{fig:rhic-pery-pTsp}
\end{figure}

\begin{figure}[tb]
\begin{center}
\includegraphics[angle=0,width=0.45 \textwidth]{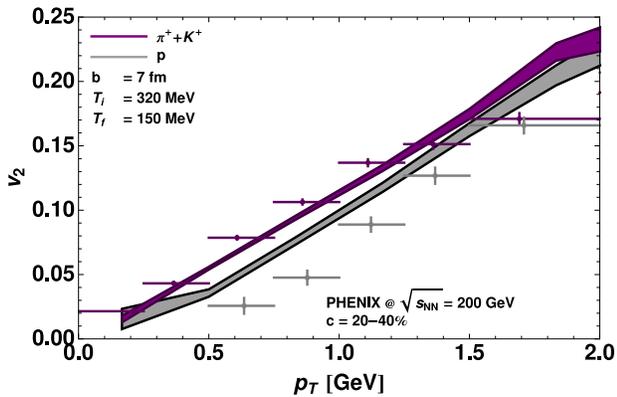}
\end{center}
\caption{(Color online) The elliptic flow coefficient $v_2$. The  values measured by PHENIX \cite{Adler:2003kt} at $\sqrt{s_{NN}}= 200~{\rm GeV}$ and the centrality class 20-40\% are indicated by the upper (pions + kaons) and lower (protons) points, with the horizontal bars indicating the $p_T$ bin. The corresponding model calculations are indicated by the solid lines, with the bands displaying 
the statistical error of the Monte-Carlo method. The parameters are the same as in Fig.~\ref{fig:rhic-pery-HS}.  }
\label{fig:rhic-pery-v2}
\end{figure}

\begin{figure}[tb]
\begin{center}
\includegraphics[angle=0,width=0.4\textwidth]{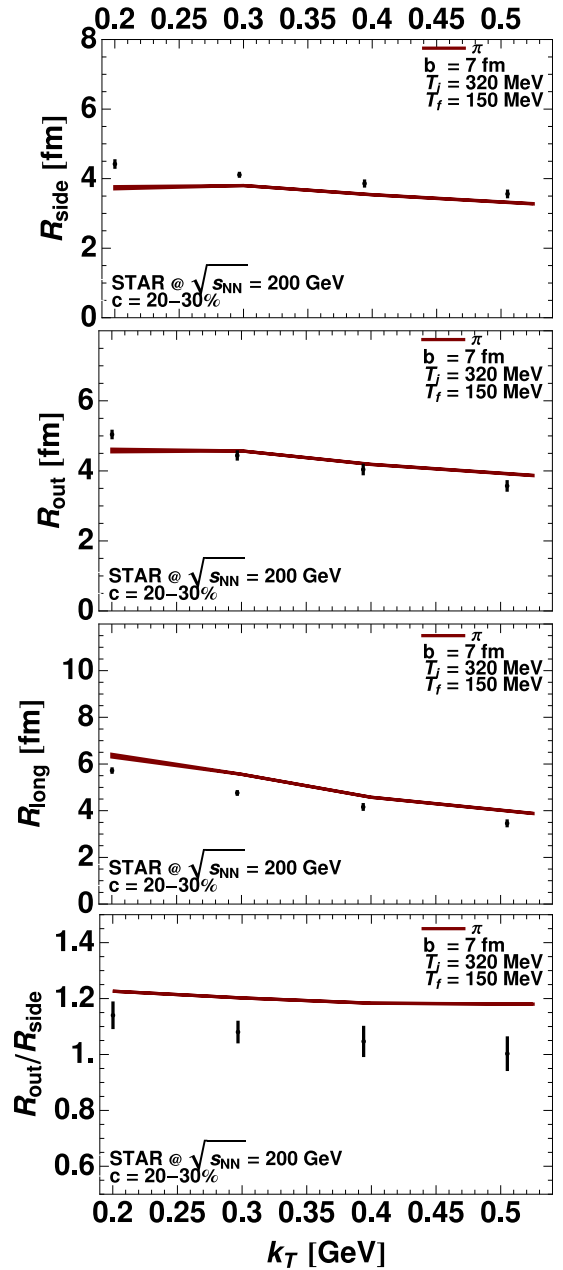}
\end{center}
\caption{The pionic HBT radii plotted as functions of the average transverse momentum of the pair compared to the STAR data \cite{Adams:2004yc} at centrality 20-30\%. As in the case of the central collisions, the calculation uses the two-particle method and includes the Coulomb effects. The values of the model parameters are the same as in Fig. \ref{fig:rhic-pery-HS}. }
\label{fig:rhic-pery-HBT}
\end{figure}

In Fig. \ref{fig:rhic-cent-HBT} the model results and the STAR data \cite{Adams:2004yc} for the HBT radii are presented. Again, for the same values of the parameters, a quite reasonable agreement is found. Discrepancies at the level of 10-15\% are observed in the behavior of the $R_{\rm side}$, which is too small, and $R_{\rm long}$, which is too large, probably due to the assumption of strict boost invariance. The ratio $R_{\rm out}/R_{\rm side}\simeq 1.25$ is larger than one, which is a typical discrepancy of hydrodynamic studies. Nevertheless, when compared to other hydrodynamic calculations, our ratio $R_{\rm out}/R_{\rm side}$ is significantly closer to the experimental value. The ratio is almost constant as a function 
of $k_T$, contrary to the decrease observed in the data.


\subsection{Non-central collisions}
\label{sec:RHIC-peripheral}

Next, we consider the centrality classes \mbox{20 - 30\%} and \mbox{20 - 40\%}. The data are compared with the model results obtained with the impact parameter \mbox{$b$ = 7 fm}. The values of the initial central temperature and the final temperature are the same as in the case of the central collisions, {\em i.e.}, $T_i = 320~{\rm MeV}$ and $T_f = 150~{\rm MeV}$. The two freeze-out curves are shown in Fig.~\ref{fig:rhic-pery-HS}. 
We observe that the out-of-plane profile is wider than the in-plane profile. This difference indicates that the system is elongated along the $y$ axis at the moment of freeze-out. This feature is in qualitative agreement with the HBT measurements of the azimuthal dependence of  $R_{\rm side}$. The comparison of the experimental and model transverse-momentum spectra is presented in Fig.~\ref{fig:rhic-pery-pTsp}. In this case we also find a good agreement between the data and the model results up to the transverse-momentum reaching 1.7~GeV.

In Fig. \ref{fig:rhic-pery-v2} the model results and the PHENIX data \cite{Adler:2003kt} on the elliptic flow coefficient $v_2(p_T)$ are compared. We observe that the $v_2$ of pions+kaons is about 10\% below the data. Taking into account the uncertainly in the initial eccentricity, discussed at the end of Sect.~\ref{sect:initcond}, which is at the level of 10-20\%, the obtained agreement is reasonable. On the other hand, the model predictions for $v_2$ of protons is too large. The discrepancy is probably caused by the final-state elastic interactions, not included in our approach \cite{Eskola:2007zc}. 


\begin{figure}[tb]
\begin{center}
\includegraphics[angle=0,width=0.4 \textwidth]{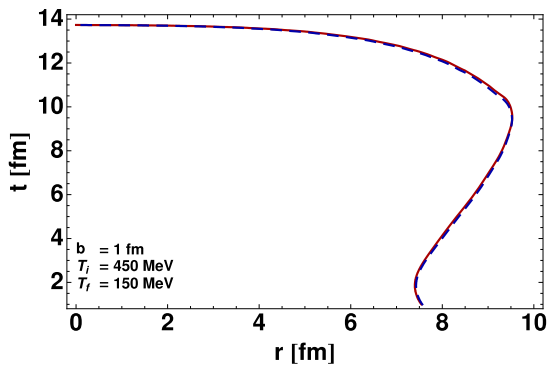}
\end{center}
\caption{(Color online) The freeze-out curves for central LHC collisions; $b = 1~{\rm fm}$,  $T_i = 450~{\rm MeV}$, and $T_f = 150~{\rm MeV}$. Similarly to the central RHIC collisions, the two curves overlap indicating that the system at freeze-out is symmetric in the transverse plane. }
\label{fig:lhc_cent-HS}
\end{figure}

\begin{figure}[tb]
\begin{center}
\includegraphics[angle=0,width=0.45 \textwidth]{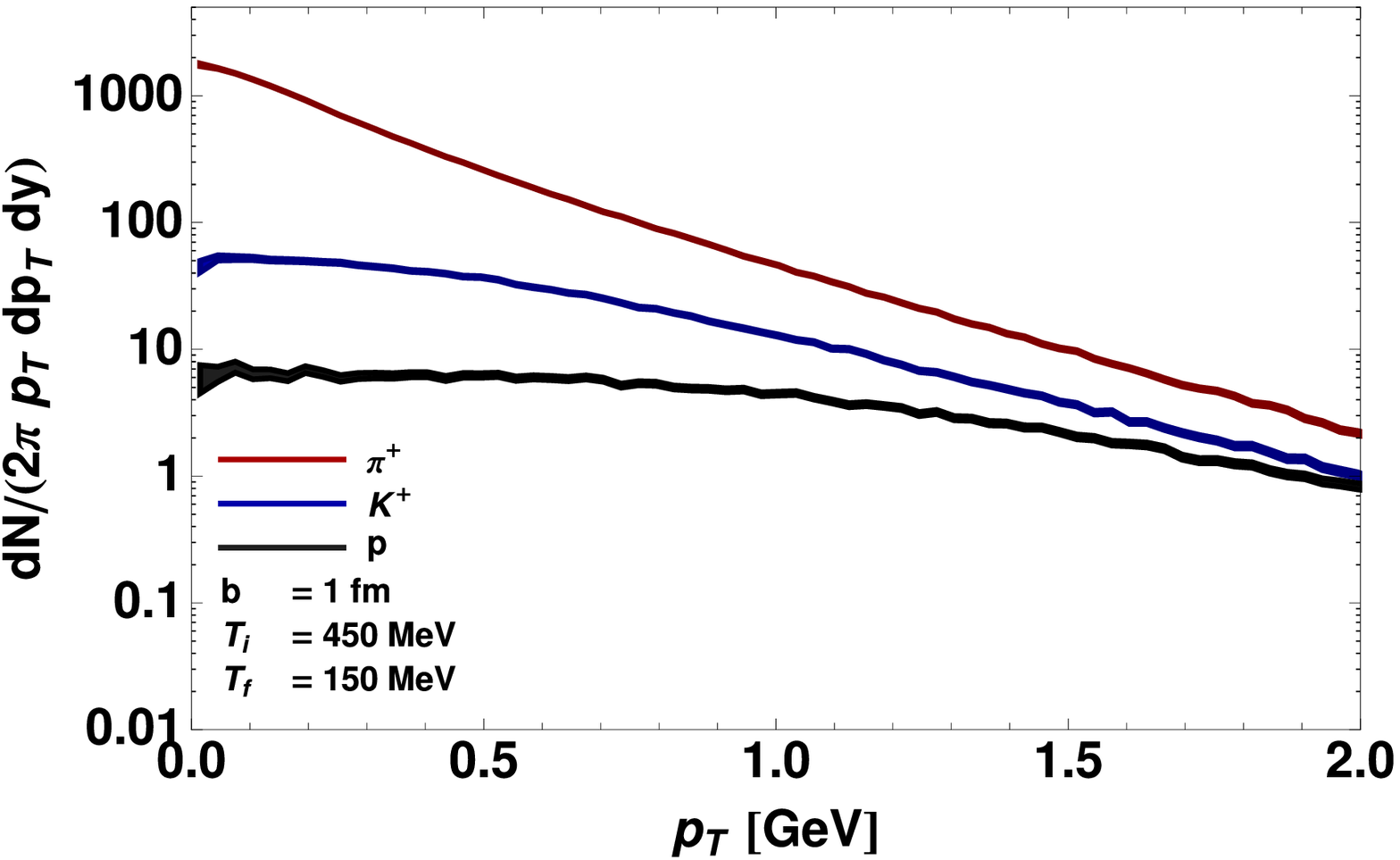}
\end{center}
\caption{(Color online) The model results for the transverse-momentum spectra of $\pi^+$, $K^+$, and protons. The values of the model parameters are the same as in Fig. \ref{fig:lhc_cent-HS}. }
\label{fig:lhc_cent-pTsp}
\end{figure}

\begin{figure}[tb]
\begin{center}
\includegraphics[angle=0,width=0.4\textwidth]{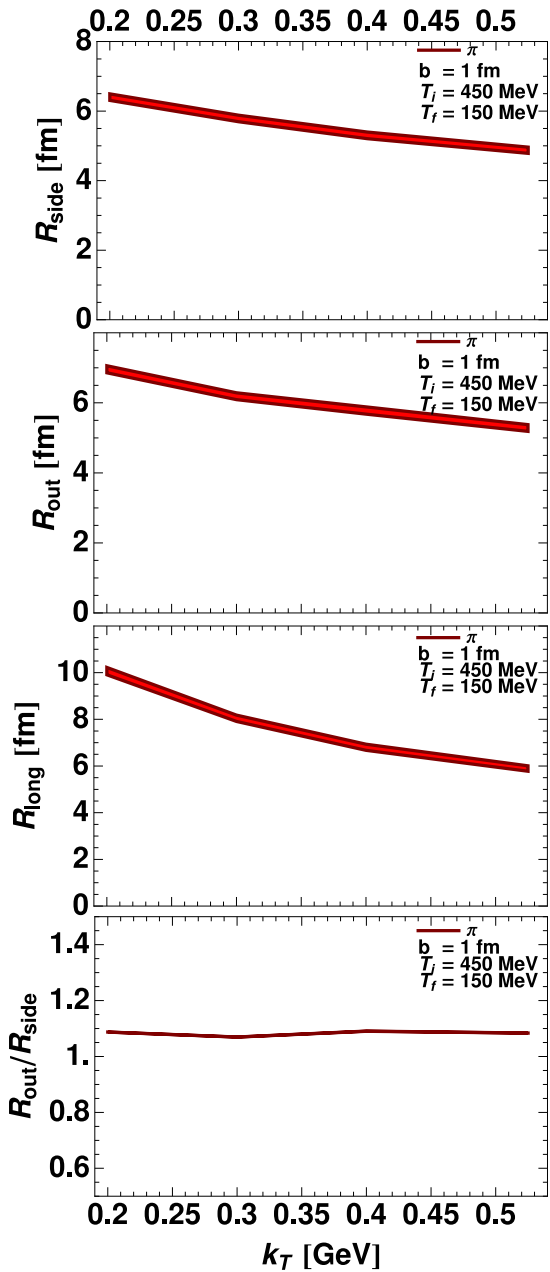}
\end{center}
\caption{The model results for the pionic HBT radii. The calculation uses the two-particle method and includes the Coulomb effects.  The values of the model parameters are the same as in Fig. \ref{fig:lhc_cent-HS}.}
\label{fig:lhc_cent-HBT}
\end{figure}

The HBT radii for non-central RHIC collisions are shown in Fig. \ref{fig:rhic-pery-HBT}. Similarly to the central collisions, we observe that $R_{\rm side}$ is slightly too small and $R_{\rm long}$ is slightly too large. Still, the ratio $R_{\rm out}/R_{\rm side}$ is very close to the data. Comparing our values of $R_{\rm side}$ with other hydrodynamic calculations we find that our values are larger. This effect is caused by the halo of decaying resonances which increases the system size by about 1 fm, see Ref.~\cite{Kisiel:2006is}.

\section{Prediction for the Large Hadron Collider}
\label{sec:LHC}

We expect that increasing the initial beam energy results essentially in a higher initial temperature $T_i$ used as the input for the hydrodynamic calculation (in the midrapidity region studied here). Therefore, to make predictions for the collisions at the LHC energies we use a set of values for $T_i$ which are higher than that used at RHIC, namely, $T_i$ = 400, 450, and 500 MeV. It is not clear which value of $T_i$ will be ``selected'' in LHC. Estimates based on extrapolations \cite{Abreu:2007kv} suggest an increase of multiplicity by a factor of 2 compared to the highest RHIC energies, which would favor $T_i$ around 400~MeV. However, to investigate a broader range of possibilities, we consider 
much higher temperatures as well. As alredy mentioned, in the case of LHC we also use a larger value of the nucleon-nucleon cross section in the definition of the initial conditions ($\sigma =66~{\rm mb}$) and different values of the chemical potentials in the hadronization made by {\tt THERMINATOR} ($\mu_B = 0.8$ and $\mu_s = \mu_{I_3} = 0$). 

In Tables~\ref{tab:lhc-cent} and \ref{tab:lhc-pery} we list our results for the following quantitites: the total $\pi^+$ multiplicity $dN/dy$, the inverse-slope parameter for pions $\lambda$, the pion+kaon elliptic flow $v_2$ at \mbox{$p_T= 1$~GeV}, and the three HBT radii calculated at the pion pair momentum $k_T= 300$~MeV. The inverse-slope parameter given in Tables~\ref{tab:lhc-cent} and \ref{tab:lhc-pery} is obtained from the formula
\begin{equation}
\lambda = - \left[ \frac{d}{dp_T} \ln 
\left( 
\frac{dN_\pi}{2\pi p_T dp_T dy}
\right)
\right]^{-1}
\label{lambda}
\end{equation}

The tables show several expected qualitative features. Obviously, as the initial temperature increases, the multiplicity grows. This is due a larger initial entropy, which causes a larger size of the freeze-out hypersurface. We find that the following parameterizations work very well for the multiplicity of $\pi^+$ at LHC for the two considered centrality cases:
\begin{eqnarray}
&&\frac{dN}{dy}=12600 (T_i/{\rm GeV})^{3.4}, \;\;\;\; (b=1~{\rm fm}) \label{dNform} \\
&&\frac{dN}{dy}=6870 (T_i/{\rm GeV})^{3.4}, \;\;\;\; (b=7~{\rm fm}). \nonumber
\end{eqnarray} 
The power 3.4 works remarkably well. This behavior reflects the dependence of the initial entropy on $T$ as shown in top left 
panel of Fig.~\ref{fig:thermo}, where for the relevant temperature range of $300-500$~MeV we have the approximate scaling 
$s/T^3 \sim T^{0.4}$. 

Similarly, for the slopes in the studied domain we have 
\begin{eqnarray}
&&\lambda(1~{\rm GeV})=0.629 T_i + 0.034~{\rm GeV}, \;\; (b=1~{\rm fm}) \label{lamform} \\
&&\lambda(1~{\rm GeV})=0.629 T_i+0.044~{\rm GeV}, \;\; (b=7~{\rm fm}). \nonumber
\end{eqnarray} 
The interesting feature is a constant ({\em i.e.} independent of $T_i$) shift by about 10~MeV from $b=1$ to $b=7$. The HBT radii increase rather moderately with $T_i$, as can be seen from the tables.

In the following Sections we show more details, discussing the spectra, $v_2$, and the HBT 
radii obtained for the initial temperature $T_i$ = 450 MeV.

\subsection{Central collisions}
\label{sec:lhc-central}

In Fig. \ref{fig:lhc_cent-HS} we show the freeze-out curves obtained from our hydrodynamic code with $T_i$ = 450 MeV. Comparing to the corresponding central RHIC collisions with \mbox{$T_i$ = 320 MeV} from Fig. \ref{fig:rhic-cent-HS}, we observe that the difference in the initial temperature results  in the longer time of the hydrodynamic expansion and a larger transverse size (both increase by about 3 fm). On the other hand, similarly to the RHIC results, we find that the two freeze-out profiles overlap, hence the system at freeze-out is, as expected, 
azimuthally symmetric in the transverse plane. We also note that the shape of the isotherms is consistent with the result presented 
in Fig. 4 of Ref. \cite{Eskola:2005ue}.

\begin{table}[t]
\begin{center}
\begin{small}
\begin{tabular}{|c|r|c|c|c|c|c|}\hline
$T_i$ [MeV] &  $\frac{dN}{dy}$ & $\lambda$ [MeV] &  $R_{\rm side}$ [fm] & 
$R_{\rm out}$ [fm] & $R_{\rm long}$ [fm]  \\ \hline 
320   &   250  &  235  &  4.3 & 5.4 & 6.1  \\ \hline 
400   &   558  &  286  &  5.3 & 6.0 & 7.6 \\ 
450   &   837  &  318  &  5.8 & 6.2 & 8.0 \\
500   &  1193  &  348  &  6.3 & 6.5 & 8.6 \\ \hline  
\end{tabular}
\end{small}
\end{center}
\caption{Central collisions at RHIC (the second row) and LHC (the three lower rows): A set of our results obtained for \mbox{$b$ = 1 fm} and four different values of the initial temperature: $T_i$ = 320, 400, 450, and 500 MeV. The columns contain the following information: $dN/dy$ -- the total pion multiplicity  (positive pions only), $\lambda$ -- the inverse-slope parameter for positive pions at \mbox{$p_T$ = 1 GeV},  \mbox{$R_{\rm side, out, long}$} -- the three HBT radii calculated at the average momentum \mbox{$k_T$ = 300 MeV.}}
\label{tab:lhc-cent}
\end{table}

\begin{table}[b]
\begin{center}
\begin{tabular}{|c|c|c|c|c|c|c|}\hline
$T_i$ [MeV] &  $\frac{dN}{dy}$ & $\lambda$ [MeV] & $v^{\pi+K}_2$ & $R_{\rm side}$ [fm] & 
$R_{\rm out}$ [fm] & $R_{\rm long}$ [fm]  \\ \hline 
320   & 131   &  245  & 0.11  & 3.8  & 4.6  & 5.6 \\ \hline
400   & 303   &  294  & 0.11  & 4.7  & 5.2  & 6.7 \\ 
450   & 455   &  327  & 0.11  & 5.1  & 5.2  & 7.1 \\ 
500   & 651   &  358  & 0.11  & 5.5  & 5.3  & 7.5 \\ \hline  
\end{tabular}
\end{center}
\caption{Non-central collisions at RHIC (the second row) and LHC (the three lower rows) ($b$ = 7 fm): The same quantities shown as in Table~\ref{tab:lhc-cent} with the additional information on the pion elliptic flow $v_2$ at \mbox{$p_T$ = 1 GeV}.}
\label{tab:lhc-pery}
\end{table}

In Fig.~\ref{fig:lhc_cent-pTsp} we give the model transverse-momentum spectra of hadrons. Compared to the RHIC results from Fig.~\ref{fig:rhic-cent-pTsp}, we find much larger multiplicities of the produced hadrons and smaller slopes of the spectra, indicating the larger transverse flow that is caused by the larger initial temperature.

Our model calculations of the HBT radii are shown in Fig.~\ref{fig:lhc_cent-HBT}. The increase of the central temperature from  \mbox{$T_i= 320$~MeV} to \mbox{$T_i = 450$~MeV} makes all the radii moderately larger. The ratio $R_{\rm out}/R_{\rm side}$ decreases by about 10\% which is an effect of the larger transverse flow caused, in turn, by the larger initial temperature.


\begin{figure}[tb]
\begin{center}
\includegraphics[angle=0,width=0.4 \textwidth]{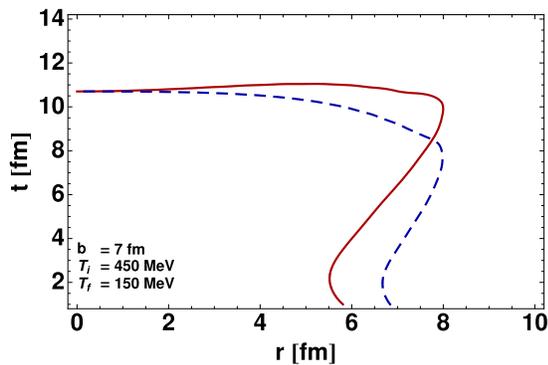}
\end{center}
\caption{(Color online) The freeze-out curves for peripheral collisions at LHC, \mbox{$b = 7~{\rm fm}$},  \mbox{$T_i = 450~{\rm MeV}$}, and \mbox{$T_f = 150~{\rm MeV}$}. The solid (dashed) line shows the in-plane (out-of-plane) profile.}
\label{fig:lhc-pery-HS}
\end{figure}

\begin{figure}[tb]
\begin{center}
\includegraphics[angle=0,width=0.45 \textwidth]{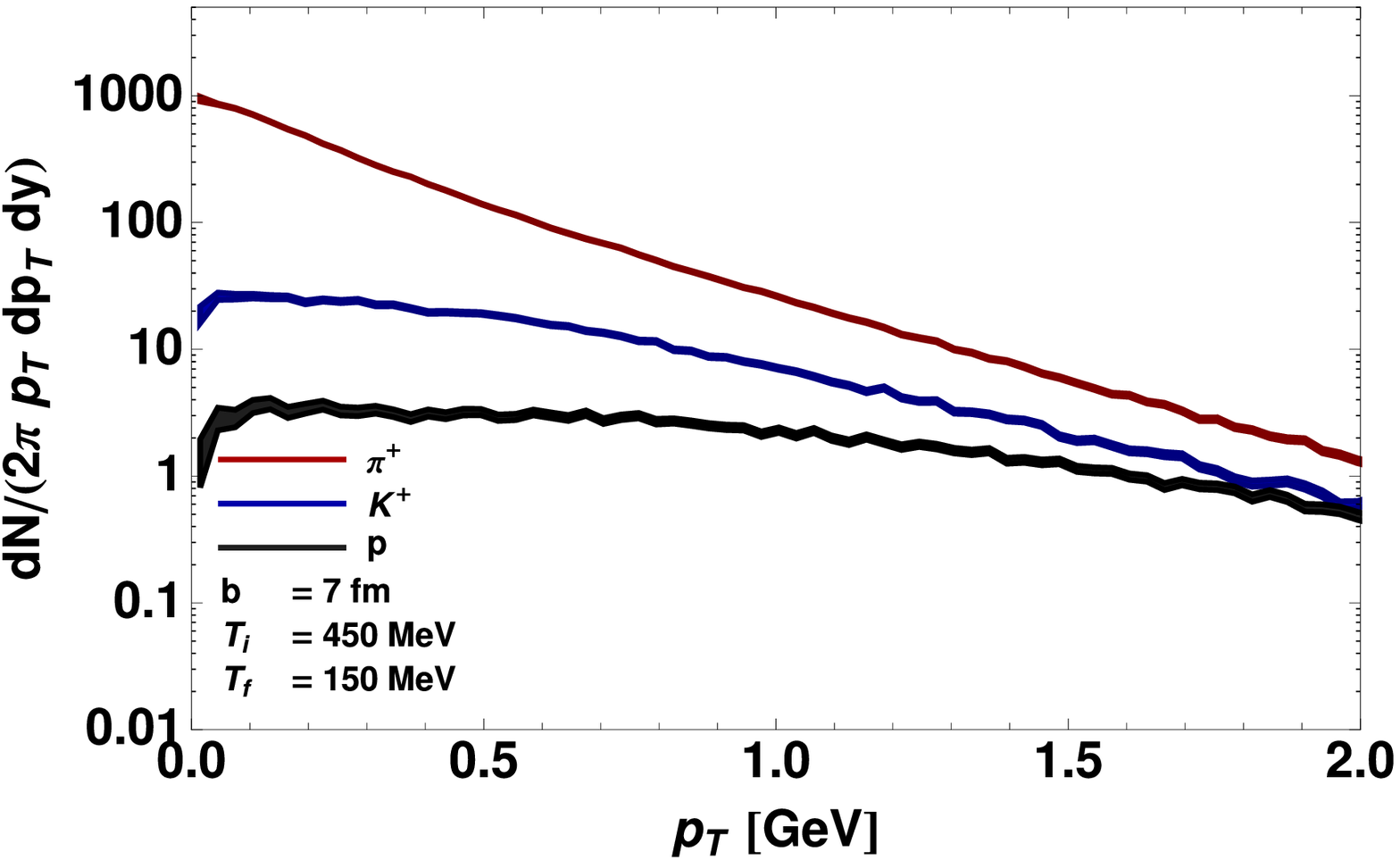}
\end{center}
\caption{(Color online) The model results for the transverse-momentum spectra of $\pi^+$, $K^+$, and protons. The values of the model parameters are the same as in Fig. \ref{fig:lhc-pery-HS}.  }
\label{fig:lhc-pery-pTsp}
\end{figure}

\begin{figure}[tb]
\begin{center}
\includegraphics[angle=0,width=0.45 \textwidth]{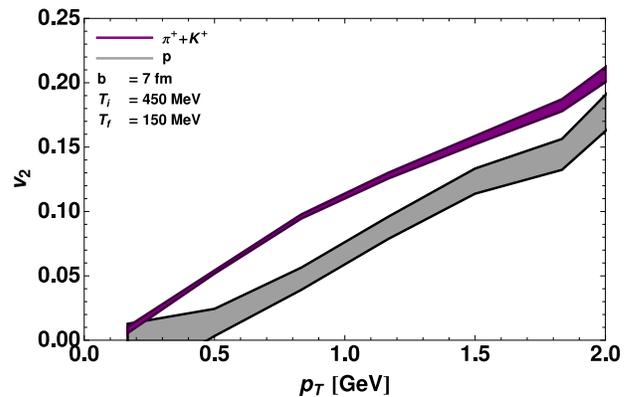}
\end{center}
\caption{(Color online) The elliptic flow coefficient $v_2$. The parameters are the same as in Fig.~\ref{fig:lhc-pery-HS}.  }
\label{fig:lhc-pery-v2}
\end{figure}

\begin{figure}[tb]
\begin{center}
\includegraphics[angle=0,width=0.4\textwidth]{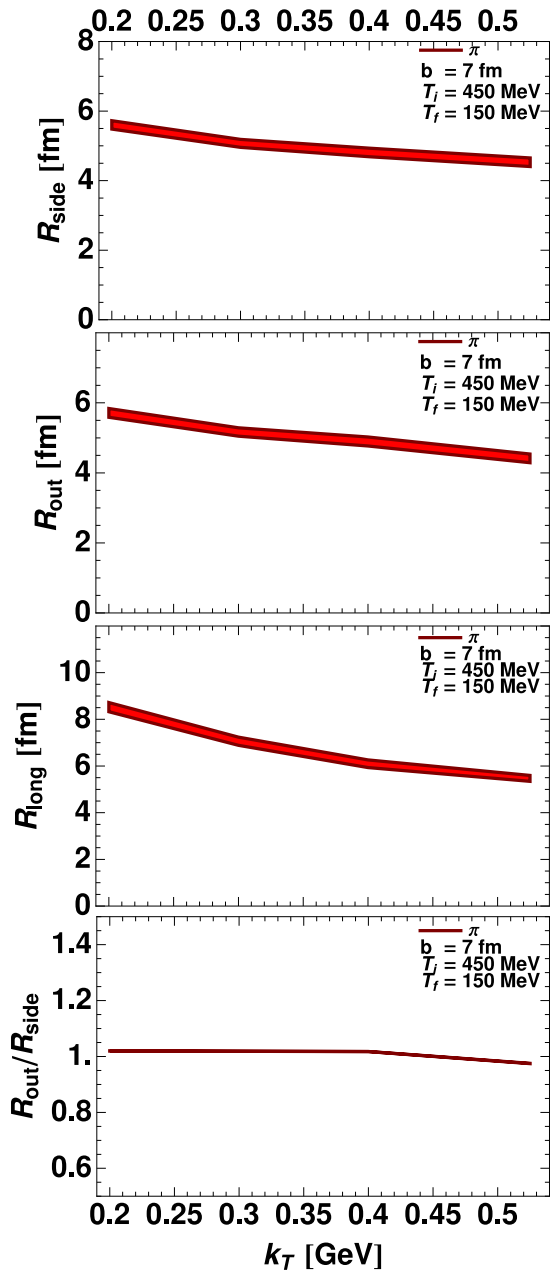}
\end{center}
\caption{The pionic HBT radii.The values of the model parameters are the same as in Fig. \ref{fig:lhc-pery-HS}. }
\label{fig:lhc-pery-HBT}
\end{figure}
 
\subsection{Non-central collisions}
\label{sec:lhc-peripheral}

In this Section we present our results describing the peripheral collisions with $T_i = 450~{\rm MeV}$, $T_f = 150~{\rm MeV}$, 
and $b = 7~{\rm fm}$. In Fig. \ref{fig:lhc-pery-HS} we show the freeze-out curves. One can observe that the system is initially elongated along the $y$ axis, but in the end of the evolution it becomes elongated along the $x$ axis. This behavior is indicated by the crossing of the freeze-out curves. The change of shape is caused by the strong flow which transforms the initial ``almond'' into a ``pumpkin'' \cite{Heinz:2002sq}. 
Azimuthally sensitive HBT probes such non-trivial behaviour and can be used as a precise confirmation test for the existence of such effects in data. 

In Fig.~\ref{fig:lhc-pery-pTsp} we show the model transverse-momentum spectra of hadrons for the same values of the parameters. Compared to the 
RHIC case with $T_i = 320~{\rm MeV}$ and the same values of $T_f$ and $b$, we find flatter spectra with higher multiplicity. 
In Fig.~\ref{fig:lhc-pery-v2} we show our results for the elliptic-flow coefficient.  The stronger transverse flow generated in this case induces larger splitting between the pion+kaon $v_2$ and the proton $v_2$, with the values of the pion+kaon elliptic flow very similar to those found in the case $T_i = 320~{\rm MeV}$. This result indicates the saturation of the elliptic flow of light particles for a given initial space asymmetry. On the other hand, the proton elliptic flow is significantly reduced. This observation is consistent with the findings of Kestin and Heinz discussed 
in Ref.~\cite{Abreu:2007kv}. Finally, in Fig.~\ref{fig:lhc-pery-HBT} we show our model calculations of the HBT radii. 
We note that the ratio $R_{\rm out}/R_{\rm side}$ is very close to one.

\section{Conclusions}

Summarizing the results, we note that the increase of the initial temperature $T_i$ of the hydrodynamic 
evolution yields a rather smooth change of the basic soft-physics observables, summarized in formulas (\ref{dNform}) - (\ref{lamform}). The observed multiplicities, spectra, and HBT radii reflect the increased values of entropy and collective flow. One may of course wonder on the credibility of the predictions, which assume that the basic picture at LHC resembles RHIC, with the initial conditions cranked up. The LHC results in the coming years will answer that question. Of course, one may 
undertake extrapolations only when the current experimental situation is described well with the given model, which is not an easy task at RHIC, where no approach describes uniformly and accurately all the available data, including the femtoscopy.    

In this paper we used the hydrodynamic approach linked to the statistical hadronization to globally describe the midrapidity hadron production at the highest RHIC energies and to make predictions for the future experiments at LHC. We have succeeded in the overall quite correct description of the soft hadron production at the highest RHIC energies, making fits of the pion spectra, $v_2$ and HBT radii. For all observables we have reached agreement at the level of 10-15\%, which should be considered satisfactory baring in mind the approximations used. We have provided uniform fits to all basic soft-physics information, including the $p_T$-spectra, the elliptic flow, and femtoscopy. We have made a best effort with the presently available tools, that is the perfect-fluid hydrodynamics with the state-of-the-art equation of state, the hadronization including all resonances, as implemented in {\tt THERMINATOR}, and the two-particle method with the Coulomb correction for evaluating the HBT radii. Having fitted RHIC at its highest energy, we have then made predictions for LHC. This was achieved by increasing the initial temperature which is, in our opinion, the main physical effect expected at larger beam energies.
Possible improvements of the present approach include the incorporation of the elastic rescattering in the final state, putting viscosity into hydrodynamics, as well as providing more accurate initial condition, including the initial evolution of partons and the fluctuations of the axes of the second harmonic of the shape generated by the Glauber Monte Carlo calculations.  

Finally, we note that {\tt THERMINATOR} may readily be used as a Monte Carlo generator 
in the energy domain of LHC also for such purposes as the detector modelling and testing. 

We are grateful to Piotr Bo\.zek for useful discussions.


\begin{thebibliography}{10}
\expandafter\ifx\csname url\endcsname\relax
  \def\url#1{\texttt{#1}}\fi
\expandafter\ifx\csname urlprefix\endcsname\relax\def\urlprefix{URL }\fi

\bibitem{Kolb:2003dz}
P.~F. Kolb, U.~Heinz, in Quark-Gluon Plasma 3, edited by R.C. Hwa and X.-N.
  Wang (World Scientific, Singapore, 2004), p. 634, nucl-th/0305084.

\bibitem{Huovinen:2003fa}
P.~Huovinen, in Quark-Gluon Plasma 3, edited by R.C. Hwa and X.-N. Wang (World
  Scientific, Singapore, 2004), p. 600, nucl-th/0305064.

\bibitem{Shuryak:2004cy}
E.~V. Shuryak, Nucl. Phys., {\bf A750} (2005) 64--83, hep-ph/0405066.

\bibitem{Muller:2007rs}
B.~Muller, arXiv:0710.3366 [nucl-th].

\bibitem{Nonaka:2007nn}
C.~Nonaka, J. Phys., {\bf G34} (2007) S313--322, nucl-th/0702082.

\bibitem{Huovinen:2001cy}
P.~Huovinen, P.~F. Kolb, U.~W. Heinz, P.~V. Ruuskanen, S.~A. Voloshin, Phys.
  Lett., {\bf B503} (2001) 58--64, hep-ph/0101136.

\bibitem{Teaney:2000cw}
D.~Teaney, J.~Lauret, E.~V. Shuryak, Phys. Rev. Lett., {\bf 86} (2001)
  4783--4786, nucl-th/0011058.

\bibitem{Teaney:2001av}
D.~Teaney, J.~Lauret, E.~V. Shuryak, nucl-th/0110037.

\bibitem{Hirano:2002ds}
T.~Hirano, K.~Tsuda, Phys. Rev., {\bf C66} (2002) 054905, nucl-th/0205043.

\bibitem{Kolb:2002ve}
P.~F. Kolb, R.~Rapp, Phys. Rev., {\bf C67} (2003) 044903, hep-ph/0210222.

\bibitem{Hama:2005dz}
Y.~Hama, {\it et~al.}, Nucl. Phys., {\bf A774} (2006) 169--178, hep-ph/0510096.

\bibitem{Eskola:2005ue}
K.~J. Eskola, H.~Honkanen, H.~Niemi, P.~V. Ruuskanen, S.~S. Rasanen, Phys.
  Rev., {\bf C72} (2005) 044904, hep-ph/0506049.

\bibitem{Nonaka:2006yn}
C.~Nonaka, S.~A. Bass, Phys. Rev., {\bf C75} (2007) 014902, nucl-th/0607018.

\bibitem{Chojnacki:2004ec}
M.~Chojnacki, W.~Florkowski, T.~Csorgo, Phys. Rev., {\bf C71} (2005) 044902,
  nucl-th/0410036.

\bibitem{Chojnacki:2006tv}
M.~Chojnacki, W.~Florkowski, Phys. Rev., {\bf C74} (2006) 034905,
  nucl-th/0603065.

\bibitem{Chojnacki:2007jc}
M.~Chojnacki, W.~Florkowski, Acta Phys. Polon., {\bf B38} (2007) 3249--3262,
  nucl-th/0702030.

\bibitem{Kisiel:2005hn}
A.~Kisiel, T.~Taluc, W.~Broniowski, W.~Florkowski, Comput. Phys. Commun., {\bf
  174} (2006) 669--687, nucl-th/0504047.

\bibitem{Abreu:2007kv}
S.~Abreu, {\it et~al.}, arXiv:0711.0974 [hep-ph].

\bibitem{Aoki:2005vt}
Y.~Aoki, Z.~Fodor, S.~D. Katz, K.~K. Szabo, JHEP, {\bf 01} (2006) 089,
  hep-lat/0510084.

\bibitem{Broniowski:2000bj}
W.~Broniowski, W.~Florkowski, Phys. Lett., {\bf B490} (2000) 223--227,
  hep-ph/0004104.

\bibitem{Bearden:2003fw}
I.~G. Bearden, {\it et~al.}, BRAHMS, Phys. Rev. Lett., {\bf 90} (2003) 102301.

\bibitem{Bearden:2004yx}
I.~G. Bearden, {\it et~al.}, BRAHMS, Phys. Rev. Lett., {\bf 94} (2005) 162301,
  nucl-ex/0403050.

\bibitem{Baym:1983sr}
G.~Baym, B.~L. Friman, J.~P. Blaizot, M.~Soyeur, W.~Czyz, Nucl. Phys., {\bf
  A407} (1983) 541--570.

\bibitem{Torrieri:2004zz}
G.~Torrieri, {\it et~al.}, Comput. Phys. Commun., {\bf 167} (2005) 229--251,
  nucl-th/0404083.

\bibitem{Amelin:2006qe}
N.~S. Amelin, {\it et~al.}, Phys. Rev., {\bf C74} (2006) 064901,
  nucl-th/0608057.

\bibitem{Florkowski:2005nh}
W.~Florkowski, Nucl. Phys., {\bf A774} (2006) 179--188, nucl-th/0509039.

\bibitem{Brown:1997ku}
D.~A. Brown, P.~Danielewicz, Phys. Lett., {\bf B398} (1997) 252--258,
  nucl-th/9701010.

\bibitem{Danielewicz:2007jn}
P.~Danielewicz, arXiv:0707.0377 [nucl-th].

\bibitem{Vertesi:2007ki}
R.~Vertesi, PHENIX, arXiv:0706.4409 [nucl-th].

\bibitem{Brown:2007raa}
D.~A. Brown, R.~Soltz, J.~Newby, A.~Kisiel, Phys. Rev., {\bf C76} (2007)
  044906, arXiv:0705.1337 [nucl-th].

\bibitem{Florkowski:2001fp}
W.~Florkowski, W.~Broniowski, M.~Michalec, Acta Phys. Polon., {\bf B33} (2002)
  761--769, nucl-th/0106009.

\bibitem{Braun-Munzinger:2001ip}
P.~Braun-Munzinger, D.~Magestro, K.~Redlich, J.~Stachel, Phys. Lett., {\bf
  B518} (2001) 41--46, hep-ph/0105229.

\bibitem{Baran:2003nm}
A.~Baran, W.~Broniowski, W.~Florkowski, Acta Phys. Polon., {\bf B35} (2004)
  779--798, nucl-th/0305075.

\bibitem{Cleymans:2004pp}
J.~Cleymans, B.~Kampfer, M.~Kaneta, S.~Wheaton, N.~Xu, Phys. Rev., {\bf C71}
  (2005) 054901, hep-ph/0409071.

\bibitem{Biedron:2006vf}
B.~Biedron, W.~Broniowski, Phys. Rev., {\bf C75} (2007) 054905,
  nucl-th/0610083.

\bibitem{Andronic:2005yp}
A.~Andronic, P.~Braun-Munzinger, J.~Stachel, Nucl. Phys., {\bf A772} (2006)
  167--199, nucl-th/0511071.

\bibitem{Huovinen:2005gy}
P.~Huovinen, Nucl. Phys., {\bf A761} (2005) 296--312, nucl-th/0505036.

\bibitem{Blaizot:1987cc}
J.~P. Blaizot, J.-Y. Ollitrault, Phys. Lett., {\bf B191} (1987) 21--26.

\bibitem{Dyrek:1984xz}
A.~Dyrek, W.~Florkowski, Acta Phys. Polon., {\bf B15} (1984) 653--666.

\bibitem{Back:2001xy}
B.~B. Back, {\it et~al.}, PHOBOS, Phys. Rev., {\bf C65} (2002) 031901,
  nucl-ex/0105011.

\bibitem{Back:2004dy}
B.~B. Back, {\it et~al.}, PHOBOS, Phys. Rev., {\bf C70} (2004) 021902,
  nucl-ex/0405027.

\bibitem{Bialas:1976ed}
A.~Bialas, M.~Bleszynski, W.~Czyz, Nucl. Phys., {\bf B111} (1976) 461.

\bibitem{Miller:2007ri}
M.~L. Miller, K.~Reygers, S.~J. Sanders, P.~Steinberg, nucl-ex/0701025.

\bibitem{Kolb:2000sd}
P.~F. Kolb, J.~Sollfrank, U.~W. Heinz, Phys. Rev., {\bf C62} (2000) 054909,
  hep-ph/0006129.

\bibitem{Aguiar:2000hw}
C.~E. Aguiar, T.~Kodama, T.~Osada, Y.~Hama, J. Phys., {\bf G27} (2001) 75--94,
  hep-ph/0006239.

\bibitem{Aguiar:2001ac}
C.~E. Aguiar, Y.~Hama, T.~Kodama, T.~Osada, Nucl. Phys., {\bf A698} (2002)
  639--642, hep-ph/0106266.

\bibitem{Socolowski:2004hw}
J.~Socolowski, O., F.~Grassi, Y.~Hama, T.~Kodama, Phys. Rev. Lett., {\bf 93}
  (2004) 182301, hep-ph/0405181.

\bibitem{Voloshin:2006gz}
S.~A. Voloshin, nucl-th/0606022.

\bibitem{Broniowski:2007ft}
W.~Broniowski, P.~Bozek, M.~Rybczynski, Phys. Rev., {\bf C76} (2007) 054905,
  arXiv:0706.4266 [nucl-th].

\bibitem{Broniowski:2007nz}
W.~Broniowski, M.~Rybczynski, P.~Bozek, arXiv:0710.5731 [nucl-th].

\bibitem{Voloshin:2007pc}
S.~A. Voloshin, A.~M. Poskanzer, A.~Tang, G.~Wang, arXiv:0708.0800 [nucl-th].

\bibitem{Alver:2007cs}
B.~Alver, {\it et~al.}, arXiv:0711.3724 [nucl-ex].

\bibitem{Cooper:1974mv}
F.~Cooper, G.~Frye, Phys. Rev., {\bf D10} (1974) 186.

\bibitem{Bugaev:2004kq}
K.~A. Bugaev, Phys. Rev., {\bf C70} (2004) 034903, nucl-th/0401060.

\bibitem{Kisiel:2006is}
A.~Kisiel, W.~Florkowski, W.~Broniowski, Phys. Rev., {\bf C73} (2006) 064902,
  nucl-th/0602039.

\bibitem{Kisiel:2006yv}
A.~Kisiel, Braz. J. Phys., {\bf 37} (2007) 917--924, nucl-th/0612052.

\bibitem{Bowler:1991vx}
M.~G. Bowler, Phys. Lett., {\bf B270} (1991) 69--74.

\bibitem{Sinyukov:1998fc}
Y.~Sinyukov, R.~Lednicky, S.~V. Akkelin, J.~Pluta, B.~Erazmus, Phys. Lett.,
  {\bf B432} (1998) 248--257.

\bibitem{Adams:2004yc}
J.~Adams, {\it et~al.}, STAR, Phys. Rev., {\bf C71} (2005) 044906,
  nucl-ex/0411036.

\bibitem{Adler:2003cb}
S.~S. Adler, {\it et~al.}, PHENIX, Phys. Rev., {\bf C69} (2004) 034909,
  nucl-ex/0307022.

\bibitem{Adler:2003kt}
S.~S. Adler, {\it et~al.}, PHENIX, Phys. Rev. Lett., {\bf 91} (2003) 182301,
  nucl-ex/0305013.

\bibitem{Broniowski:2001ei}
W.~Broniowski, W.~Florkowski, Phys. Rev., {\bf C65} (2002) 024905,
  nucl-th/0110020.

\bibitem{Broniowski:2001we}
W.~Broniowski, W.~Florkowski, Phys. Rev. Lett., {\bf 87} (2001) 272302,
  nucl-th/0106050.

\bibitem{Broniowski:2001uk}
W.~Broniowski, W.~Florkowski, Phys. Rev., {\bf C65} (2002) 064905,
  nucl-th/0112043.

\bibitem{Rafelski:2000by}
J.~Rafelski, J.~Letessier, Phys. Rev. Lett., {\bf 85} (2000) 4695--4698,
  hep-ph/0006200.

\bibitem{Broniowski:2002nf}
W.~Broniowski, A.~Baran, W.~Florkowski, Acta Phys. Polon., {\bf B33} (2002)
  4235--4258, hep-ph/0209286.

\bibitem{Prorok:2007xp}
D.~Prorok, nucl-th/0702042.

\bibitem{Eskola:2007zc}
K.~J. Eskola, H.~Niemi, P.~V. Ruuskanen, arXiv:0710.4476 [hep-ph].

\bibitem{Heinz:2002sq}
U.~W. Heinz, P.~F. Kolb, Phys. Lett., {\bf B542} (2002) 216--222,
  hep-ph/0206278.

\end{thebibliography}

\end{document}